\documentclass[pre,twocolumn,superscriptaddress,showpacs]{revtex4}

\usepackage{graphicx,color}
\usepackage{amsmath}
\definecolor{r}{rgb}{1,0,0}
\definecolor{b}{rgb}{0,0,1}

\begin{document}

\newcommand{\LANL}{Condensed Matter and Thermal Physics and Center
for Nonlinear Studies, Los Alamos National Lab, NM, 87545, USA}

\newcommand{\SZFKI}{Research Institute for Solid State Physics and Optics,
                       POB 49, H-1525 Budapest, Hungary}

\title{Flow rule of dense granular flows down a rough incline}

\author{Tam\'as B\"orzs\"onyi}
\email{btamas@szfki.hu}
     \affiliation{\LANL}
     \affiliation{\SZFKI}

\author{Robert E. Ecke}
     \affiliation{\LANL}

\date{\today}

\begin{abstract}
We present experimental findings on the flow rule for granular flows on a
rough inclined plane using various materials including sand and glass beads
of various sizes and four types of copper particles with different shapes.
We characterize the materials by measuring $h_s$ (the thickness at which
the flow subsides) as a function of the plane inclination $\theta$ on
various surfaces.
Measuring the surface velocity $u$ of the flow as a function of flow
thickness $h$, we find  that for sand and glass beads the Pouliquen flow rule
$u/\sqrt{gh} \sim \beta\ h/h_s$  provides reasonable but not perfect
collapse of the $u(h)$ curves measured for various $\theta$ and mean 
particle diameter $d$.
Improved collapse is obtained for sand and glass beads by using a recently
proposed scaling of the form $u/\sqrt{gh} =
\beta \cdot h\ \tan^2\theta /h_s\ \tan^2\theta_1$ where $\theta_1$
is the angle at which the $h_s(\theta)$ curves diverge.
Measuring the slope $\beta$ for ten different sizes of sand and glass beads, we
find a systematic, strong increase of $\beta$ with the divergence 
angle $\theta_1$
of $h_s$. The copper materials with different shapes are not well described
by either flow rule with $u \sim h^{3/2}$.
\end{abstract}

\pacs{47.57.Gc, 45.70.-n}

\maketitle

\section{Introduction}
\label{intro}

Granular flow on a rough inclined plane is an important system with which
to learn about the basic rules of the dynamics of granular materials
\cite{ba1954,humo1984,po1999,fopo2003,azch1999,an2001,ra2003,bede2003,ra2005}.
Despite intensive study, the fundamental features of such flows are still
incompletely understood (for reviews see: 
\cite{jana1996,gdrmidi2004,arts2006}).
The majority of laboratory experiments report on the flow properties in narrow
channels (quasi-2D geometry) where the velocity can be measured as a function
of depth by directly viewing grain motion through the sidewalls
\cite{azch1999,ra2003,koin2001,an2001,bede2003,bide2005,ra2004,ra2005,gdrmidi2004}.
In this configuration, however, the effect of friction with the confining
vertical walls is important \cite{khor2001,bide2005,gdrmidi2004,tari2003},
and remains a determining force for thick flows (flow on a pile) even in
wider channels \cite{tari2003,jofo2005}.

For thin flows in wide channels measuring the depth dependence of the flow
velocity (far from the side walls) is far more difficult. To characterize
the basic features of granular flows in this configuration, the surface
velocity $u$ or the depth averaged velocity $U$ can be measured as a function
of the flow thickness $h$. The depth averaged flow velocity $U$, inferred from
the front velocity of the granular layer, was
systematically measured by Pouliquen as a function of the flow 
thickness $h$ for glass
beads over a range of plane inclinations $\theta$ \cite{po1999}.  
The $U(h)$ curves measured at different values of $\theta$ collapsed 
when using the scaling
law $U/\sqrt{gh}=\beta\ h/h_s - \gamma$ (where $h_s$ corresponds to the
thickness where the flow subsides) giving rise to a general flow 
rule, denoted the
``Pouliquen flow rule,'' for
glass beads with various sizes and for which $\beta \approx 0.14$ and
$\gamma \approx 0$. It was subsequently reported \cite{fopo2003} that the
same scaling collapsed the $U(h)$ curves for sand with one particular size
of $d=0.8$ mm. The slope for the sand data, $\beta \approx 0.65$, was
considerably larger than for glass beads and $\gamma \approx 0.77$.
This quantitative difference in the flow rule was used to explain complex 
dynamical phenomena, such as waves \cite{fopo2003} or avalanche propagation 
\cite{boha2005}.

It is of considerable interest to determine the robustness of the Pouliquen 
flow rule (PFR) for different flow conditions including particle diameter, 
relative surface roughness, and particle shape. A further consideration is 
whether the flow rule is sensitive to measuring the front velocity as compared
to measuring the surface velocity. The former yields a better depth averaged 
velocity but is subject to saltating grains for faster flows which limited 
the accuracy of the measurement to about 10 \% \cite{po1999} and would not
be applicable for general granular materials subject to a fingering 
instability \cite{pode1997}.  The surface velocity measurement is 
characteristic of the steady flow even for general granular media and avoids
the accuracy limitations imposed by saltation, but can only be related to a 
depth-average by some assumption of the vertical velocity profile.  
Neither approach is ideal, being rather complementary as opposed to one 
being {\it a priori} better than the other.  Our work establishes the utility
and robustness of using the surface velocity to determine the flow rule.  
If a flow rule is to be a useful measure of the state of granular flow on
an incline, it should not be particularly sensitive to the details of its 
determination.

A recent theory by Jenkins \cite{je2006} suggests a phenomenological
modification of the hydrodynamic equations for dense flows. According to the
theory enduring contacts between grains forced by the shearing reduce the
collisional rate of dissipation while continuing to transmit force and momentum.
This assumption has several consequences, one of which is a modification of the
Pouliquen scaling law by the inclusion of a $\tan^2\theta$ correction to the
$h/h_s$ term.  Replotting the Pouliquen data, Jenkins found a better collapse
of the data using his modified form, denoted here as the ``Pouliquen-Jenkins'' 
flow rule (PJFR). The improvement of the collapse, however, was not definitive
owing to the scatter in the velocity data and in the associated determination 
of the $h_s(\theta)$.

One of the purposes of a flow rule is to have a compact description of easily
measurable quantities that represents the subtle balances of stress and 
strain rate in a granular material, {\it i.e.}, the granular rheology.  
Although the vertical velocity profile in a flowing granular layer has not 
been obtained experimentally, let alone the experimental determination
of local stresses and strain rates, a general discussion about possible flow
rheologies helps set a background for presenting empirical flow rules determined
from experiment. In particular, the scaling of the velocity with layer 
thickness can be understood by a consideration of bulk Bagnold rheology 
\cite{ba1954,sier2001}. In the theory of Bagnold, the shear stress varies 
with the shear rate $\dot{\gamma}$ like $\sigma_{xz} \sim \dot{\gamma}^2$. 
This relationship is based on the following assumptions.  The transport of 
the $x$ component of momentum in the $z$ direction occurs through collisions 
whose rate depends on the velocity gradient $\dot{\gamma}$. Similarly the 
momentum transfer per collision scales linearly with $\dot{\gamma}$ leading 
to the quadratic dependence between stress and strain.  With a linear 
dependence of shear stress on the vertical coordinate $z$, this leads
to a vertical variation of the down-plane velocity of
$u(z) \sim h^{3/2} \left [ 1-\left (\left (h-z\right )/h\right 
)^{3/2}\right ]$.
Thus, the surface velocity $u = u(h) \sim h^{3/2}$ so that the 
scaling $u/\sqrt{gh}$ versus $h$ (suitably corrected for inclination angle) 
should yield straight lines with zero intercept. Such scaling was reported 
for experiments \cite{po1999,dela2006} using glass spheres and for numerical
simulations of idealized spherical particles \cite{sier2001}. Also, deviations
from this law towards a linear velocity profile were reported in experimental
\cite{dela2006} and numerical \cite{sila2003} studies for thin flows. 
For a particular flow profile, the surface velocity $u$ and the depth 
averaged velocity $U$ are related by a constant factor. Thus, there is no a 
priori reason to prefer one over the other. Although the interior velocity 
profile far from sidewalls has not been measured to our knowledge, the scaling
$u \sim h^{3/2}$ is indirect support for the Bagnold flow rheology. The degree
to which such scaling fails, therefore, would appear to call for modification 
of the assumptions leading to the Bagnold rheology. We will see in this paper
how well the Bagnold-based rheology applies to a range of different granular
materials.

In the present work we investigate the flow properties of 14 different 
materials by measuring the surface velocity $u$ as a function of flow 
thickness $h$. Because of our measurement methods, the statistical uncertainty
in our data is considerably less than in previous studies \cite{po1999},
allowing for a more detailed and quantitative evaluation of different flow
rule scalings.  We find that scaling the surface flow velocity by $\sqrt{gh}$
and the flow thickness by $h_s$ assuming the Pouliquen flow rule provides
reasonable but not perfectly accurate collapse of the $u(h)$ curves taken
at various plane inclinations measured in a wider range of the main control
parameters of grain size, plane inclination angle, surface roughness and
flow thickness compared to earlier studies \cite{po1999,fopo2003,dela2006}.
Improved collapse is obtained for sand and glass beads using the modified
Pouliquen-Jenkins scaling law 
$u/\sqrt{gh} = \beta \cdot h \ \tan^2\theta /h_s \tan^2\theta_1$
where the factor $\tan^2\theta$ is supported by a recent theory \cite{je2006}.
For glass beads the straight lines of the scaled curves support the Bagnold
rheology.  For sand, although the data are well collapsed by the scaling, the 
curves are slightly concave downward suggesting high-order corrections in $h$
beyond the simple Bagnold result. We show that the slope $\beta$ of the master
curve for the sand/glass-bead materials (obtained for each material) strongly
increases with $\tan \theta_1$ or $\tan \theta_r$ where $\theta_1$ and $\theta_r$
are the angles where $h_s(\theta)$ diverges and the bulk angle of repose,
respectively. The similarities and differences of our experimental approach
compared to other experimental measurements of flow rules \cite{po1999,dela2006}
are discussed in detail.

In contrast to the relatively simple and understandable data obtained for
glass beads and for sand, the behavior of flowing copper particles is more 
complex and a simple Bagnold interpretation works quite poorly in describing
the relationship between surface velocity and layer height. Indeed, the the
scaling of $u$ with $h$ is closer to $u \sim h^{1/2}$ than to the Bagnold form
$u \sim h^{3/2}$.  Nevertheless, the angle correction using $h_s(\theta)$ 
(Pouliquen flow rule) or $h_s(\theta)/\tan^2\theta$ (Pouliquen-Jenkins flow
rule) appears to work pretty well with the latter again providing better
overall data collapse.

\section{Experiment}
\label{exp}

\begin{figure}[ht]
\resizebox{85mm}{!}{
\includegraphics{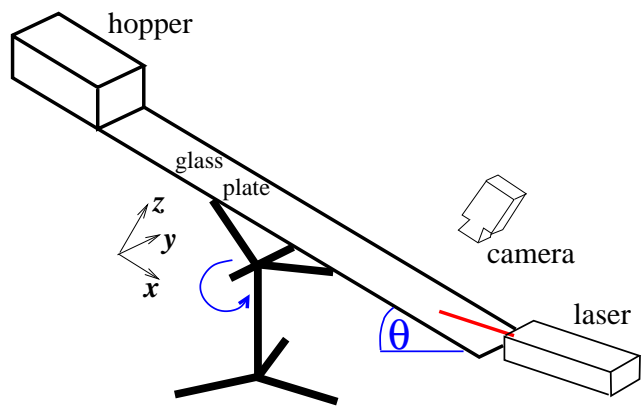}
}
\caption{
(Color online)
Schematic of the experimental setup used to measure the surface velocity
and height.  The whole system could be rotated (together with laser and 
camera) to set an arbitrary inclination angle
$\theta$.
}
\label{setup}
\end{figure}

The experimental measurements presented in this paper were performed in two
different setups. The first apparatus was described in detail elsewhere
\cite{boec2006vacuum} and consisted of a glass plate with dimensions
230 cm x 15 cm (see Fig.\ \ref{setup}). The leftmost 40 cm of the plate
served as the bottom of the hopper. The surface of the remaining part (190 cm)
of the glass plate was typically covered with sandpaper that was glued to the
surface and had a roughness of $R = 0.19$ mm (grit 80) which provided an 
extremely durable uniform roughness to the plate. Other values of plate 
roughness were studied using different grit sandpaper (and a few measurements
with 0.4 mm sand glued to the glass plate) to explore the systematic 
dependence of our results on relative roughness compared to grain size. 
The plate together with the hopper could be tilted, enabling us to set
an arbitrary inclination angle $\theta$. The flow was characterized 
by measuring the surface velocity $u$ as a function of the thickness $h$ in 
the stationary dense flow regime at a location $x_o = 155$ cm below the hopper
gate, sufficiently far downstream to have established a steady state 
\cite{boec2006vacuum}.  This first apparatus could be tilted back and forth
to recharge the hopper, facilitating the accumulation of the large amounts of
data reported here. Because the system was closed in a cylindrical tube, 
precise measurements of $h_s(\theta)$ were difficult and were performed in a 
second apparatus.

The second apparatus used to measure  $h_s(\theta)$ consisted of a wider plate
having dimensions  227 cm x 40 cm and was covered with the same sandpaper used
to cover the surface of the flow channel. The system was not confined from 
the top.  This wider, non-enclosed channel allowed for very precise 
measurements of the layer height in a rapid manner. The procedure was to throw
the grains onto the plane and allow a uniform layer thickness to form by 
letting the flow subside.  The granular material was swept from a 2 m long 
area and its volume was measured accurately, yielding a very precise and 
repeatable measurement of the mean layer thickness $h_s(\theta)$. This method
averages out the spatial variations in  $h_s$, the amplitude of which was also
estimated by measuring the displacement of a projected laser sheet. At lower 
$\theta$, the height variations were typically less than $\pm 5\%$ of $h_s$, 
but became larger at higher plane inclinations where $h_s$ became less than 
$5d$. Because the majority of the data on the flow properties were measured
at plane inclinations corresponding to these relatively lower values of $h_s$,
it is important to get an accurate measure of $h_s$. The repeatability of 
the measurement also depended on the plane inclination, but in this case 
relative variations decreased with increasing $\theta$.  The data points fell 
onto the same curve within an error of $\pm 5 \%$ for $\tan\theta/\tan\theta_r>1.1$.
When approaching $\theta_r$ the measurements became less accurate with the rapid
increase of $h_s$ leading to a $\pm 12 \%$ uncertainty of the data points for
$\tan\theta/\tan\theta_r<1.1$.

Uncertainties arising from slightly nonuniform thickness near the walls were
also estimated. We observed a boundary layer $W$ where the layer thickness was 
slightly larger than elsewhere. The width of the boundary layer was $W<1$ cm
for $\tan\theta/\tan\theta_r>1.1$ and somewhat larger $W<5$ cm for smaller $\theta$.
The effect of the boundary layer results in a slight overestimation of $h_s$
corresponding to about  $2\%$ for $\tan\theta/\tan\theta_r>1.1$ and $5\%$ for
smaller $\theta$.  To reduce the effect of the boundary layer we removed part of
the excess material near the boundary, and we estimate that the finally measured
value of $h_s$ is overestimated by less than $1\%$ owing to the effect of the
lateral boundaries.

A possible concern regarding using one apparatus to measure $h_s(\theta)$
and another to measure $u$ and $h$ in the flowing state is that the lateral
boundary effects might be different, leading to possible discrepancies in
the measurements. To that end we measured the flow profile in the narrow
channel, as presented below, and found that the flow was uniform over the
central 80\% of the narrow channel. Because our measurements of $u$ and
$h$ were taken in the center of the narrow channel, we conclude that no
significant differences arise from using different channels for the static
and dynamics measurements, respectively. Further, because of the limitations
of each system, the amount of data we obtained would not have been feasible
using one or the other of our experimental setups.

Four types of granular materials were used. The first set consisted of sand particles
from the same origin but sorted into four different sizes. For example, the finest
sample was obtained by sifting the sand with 100 and 300 $\mu$m sieves. We designate
this distribution as having a mean of $d=0.2$ mm and a standard deviation of $0.05$ mm.
According to this notation the four sets of sand correspond to sizes
$d=0.2\pm0.05$ mm,  $d=0.4\pm0.05$ mm, $d=0.6\pm0.05$ mm and $d=0.85\pm0.08$ mm
while the mean particle density was $\rho_{sand} = 2.6$ g/cm$^3$.
The fifth sample of sand originated from the Kelso dunes and was well sorted
with a size distribution of $d=0.2\pm0.05$ mm. This sand is peculiar in that
it emits sound when sheared. The Kelso dune is known to be an example
of ``booming sand dunes'' \cite{doma2006}.
\begin{figure}[ht]
\resizebox{85mm}{!}{
\includegraphics{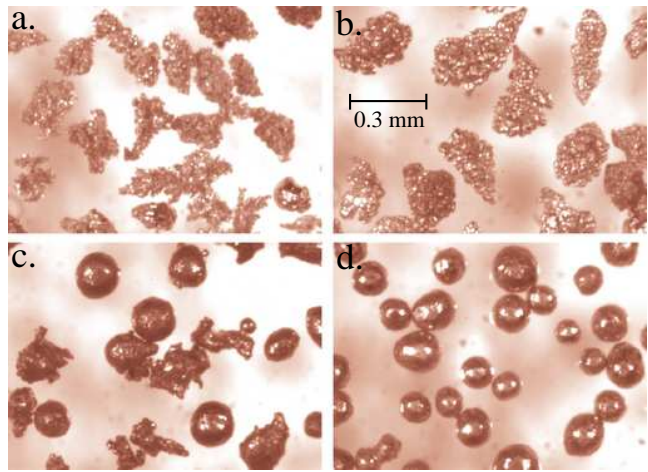}
}
\caption{
(Color online) Microscopic images of the copper particles with
$d=160\pm50$ $\mu$m and with packing fractions $\eta$:
a)  $0.25$,
b)  $0.33$,
c)  $0.5$
and d)  $0.63$.
}
\label{copper-images}
\end{figure}
We also used commercial glass beads (Cathapote) with sizes $d=0.18\pm0.05$ mm,
$d=0.36\pm0.05$ mm, $d=0.51\pm0.05$ mm and $d=0.72\pm0.08$ mm and mean particle
density of $\rho_{glass} = 2.4$ g/cm$^3$.
One sample of the $d=0.51\pm0.05$ mm glass beads was carefully washed.
For that sample, we observed a slight change in the flow properties as well
as in the value of $h_s$ compared to an unwashed sample with the same $d$
and, thus, we report these data as an additional case. The last type of material
consisted of copper particles with a mean size of $d=0.16\pm0.03$ mm but with
different shapes.  The shape anisotropy of the four different samples of copper
particles is characterized by the volume fraction $\eta$ (the ratio of the volume
occupied by the particles and the total volume) of the material at rest, with values
0.63, 0.5, 0.33, 0.25 and particle densities 8.7, 8.2, 7.6, 7.1 g/cm$^3$, respectively. 
The variation in $\eta$ represents the strong change in the shape
from spherical particles to very dendritic shapes with decreasing particle densities
for the more dendritic shapes as well.  Images of copper particles are shown in
Figs.\ \ref{copper-images}(a)-\ref{copper-images}(d) where the strong variation
in particle shapes is clearly seen.

\begin{figure}[h]
\resizebox{86mm}{!}{
\includegraphics{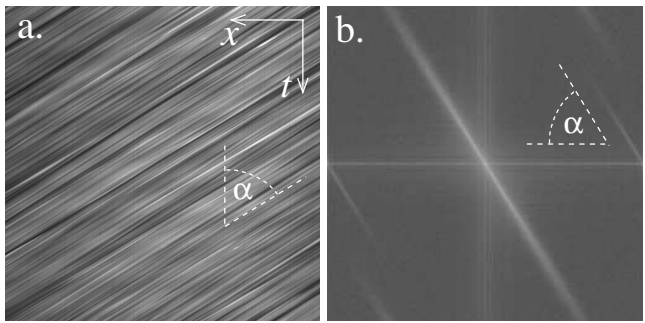}
}
\caption{
a) Space time plot showing particle streaks along a line oriented with
the flow direction for $\theta = 36.1^\circ$ and $H = 2$ cm. Dimensions
of the image are $3.68$ cm and $0.08$ s.
b) Two dimensional FFT in frequency/wave-number space of the image in
a) which gives an accurate measure of the mean flow velocity as indicated
by the solid line. The angle $\alpha$ is designated in each image.}
\label{spatio-fourier}
\end{figure}

The surface flow velocity $u$ was determined by analyzing high speed
(8000 frames per second) video recordings. Individual particles make streaks
in a space-time plot of intensity along one line of camera pixels aligned
with the mean flow direction.  An example of such a space-time image is
shown in Fig.\ \ref{spatio-fourier}(b) where the length of the line in the camera
is $L = 3.68$ cm and the total time is $T = 0.080$ s.
The streaks are generally oriented at some angle $\alpha$ in the 
image. Performing
a fast Fourier transform (FFT) produces a line perpendicular to the 
streaks, see
Fig.\ \ref{spatio-fourier}, which gives a measure of the mean surface velocity
$u=(L/T) \tan \alpha$. The thickness $h$ of the flow was monitored by 
the translation
of a laser spot that was projected onto the surface of the plane at an angle of
$\phi = 20^\circ$ in the $xz$ plane (see Fig.\ \ref{setup}). Other 
details regarding
the measurement techniques can be found in \cite{boec2006vacuum}.

\section{Results and Discussion}
\label{results}

The two measurements that determine the flow rule are the height of the
layer when the flow stops $h_s(\theta)$ and the dependence of the surface velocity
on the layer height $h$.  We first consider $h_s(\theta)$ for glass beads, sand and copper.
We then present measurements of $u$ as a function of $h$ for sand and glass beads and
the application of the flow rules of PFR and PJFR. Finally,
we consider velocity data and flow rules for the copper material.

\subsection{Determination of $h_s(\theta)$}
\begin{figure}[h]
\resizebox{74mm}{!}{
\includegraphics{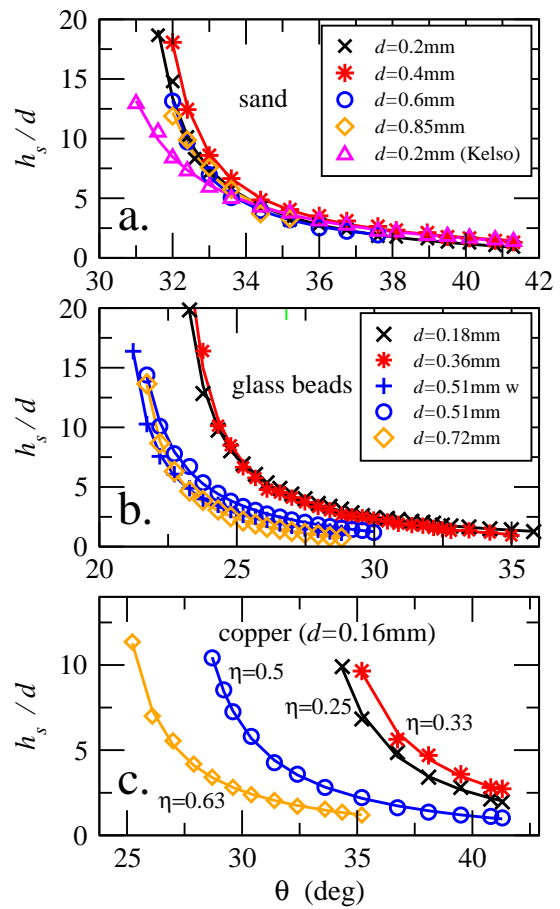}
}
\caption{
(Color online)
The thickness $h_s$ at which the flow subsides normalized by
the grain diameter $d$ as a function of the plane inclination angle $\theta$
for a) sand and b) glass beads of various sizes and c) copper particles of various
shapes (as indicated by the static volume fraction $\eta$). The grain
diameter $d$ is indicated and w designates the case of washed glass beads.
The continuous lines are best fits according to the formula
$h_s/d = B (\tan \theta_2 - \tan \theta)/(\tan \theta - \tan \theta_1)$.
The resulting values of $\theta_1$ are indicated for each material in
Table \ref{param-table}.
}
\label{staticlayer}
\end{figure}

As seen in Figs.\ \ref{staticlayer}(a)-\ref{staticlayer}(c), $h_s$ increases
rapidly with decreasing $\theta$ and diverges at $\theta_1$.
The solid lines in Fig.\ \ref{staticlayer} are best fits to the formula
$h_s/d = A (\tan\theta_2 - \tan\theta)/(\tan\theta - \tan\theta_1)$, a simple
function which diverges at $\theta_1$ and goes to zero at $\theta = \theta_2$
\cite{po1999,pofo2002,fopo2003,gdrmidi2004}.
The resulting values for the fitting parameters $A, \theta_1$, and $\theta_2$
are indicated for each material in Table \ref{param-table}.
The bulk angle of repose $\theta_r$ was also measured for several materials by
measuring the dynamics of a three dimensional sandpile under constant flux conditions.
As material was added at a very small but uniform rate to the top of the pile,
avalanches formed and propagated downward intermittently. The distribution of
the angle, observed directly after the avalanche stopped, was measured for hundreds
of avalanches. The mean of this distribution was taken to be $\theta_r$, the bulk
angle of repose.  The value of $\theta_r$ is very close to $\theta_1$
as indicated in Table \ref{param-table}.

The $h_s(\theta)$ curves are very similar for all four sand samples
originating from the same source [see Fig.\ \ref{staticlayer}(a)].
The fifth curve corresponding to the Kelso sand showed deviations from the
other data at lower values of $\theta$, yielding a somewhat smaller value
for $\theta_1$. This difference is attributable to the more rounded shape
for the Kelso sand  as revealed in microscope images.
The $h_s(\theta)$ curves for glass beads, however, formed two groups.
Microscope images revealed that the two samples with smaller $d$ contained a
larger amount of non-spherical particles than the two sets with larger $d$.

\begin{table}[h]
\resizebox{80mm}{!}{
\includegraphics{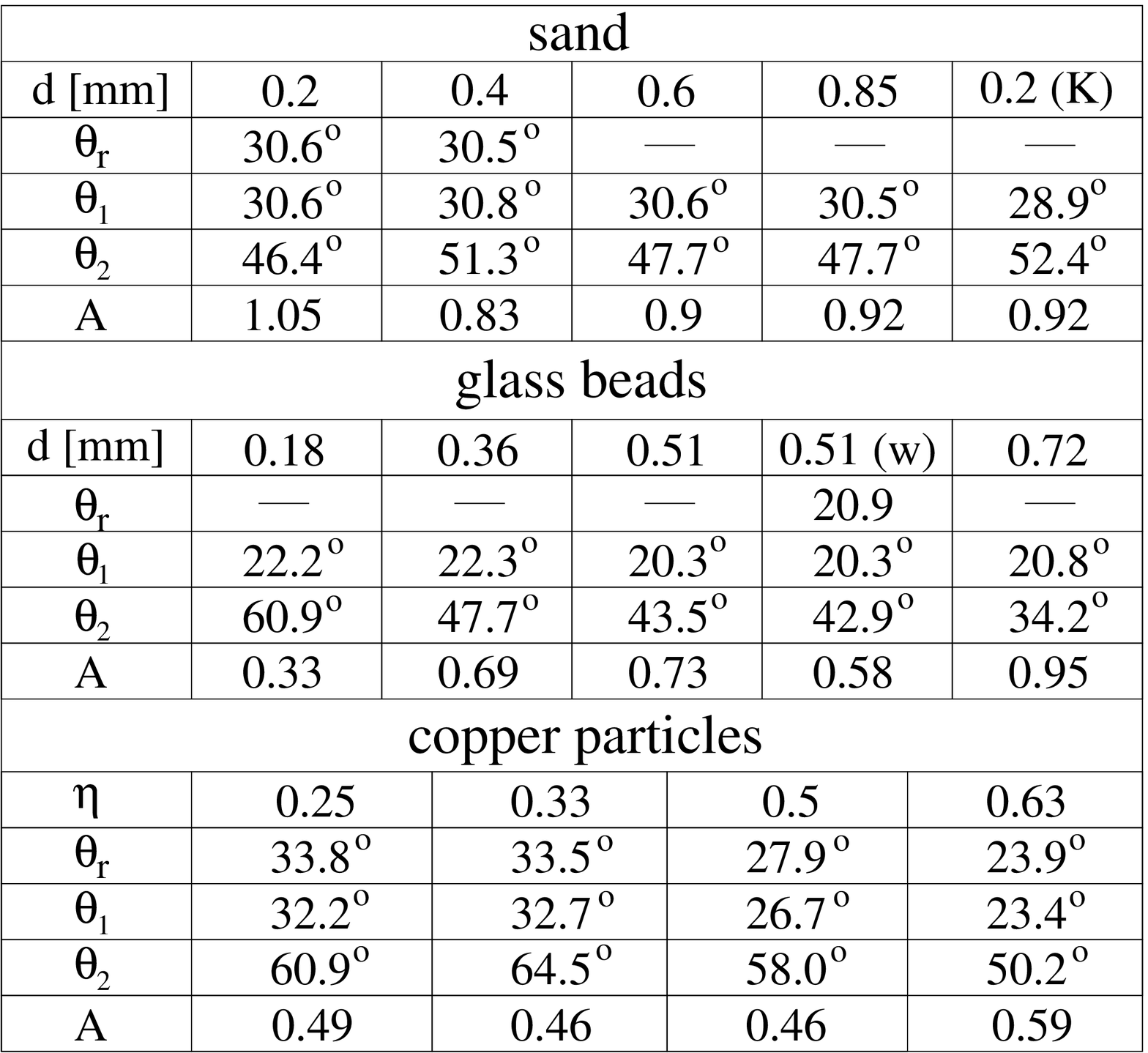}
}
\caption{
The values of  $\theta_r$, $\theta_1$, $\theta_2$ and $A$ for sand and glass
beads of size $d$ and copper particles with $d=0.16$ mm and volume fractions
$\eta$. K stands for the Kelso sand, and w denotes the washed glass beads.
}
\label{param-table}
\end{table}

The difference in shape may explain the slightly larger values of
$\theta_1$ and $h_s/d$ for the two samples with smaller $d$. The case of
the $0.51$ mm glass beads is also interesting in that washing the material
with tap water resulted in a slightly smaller value of $h_s$, which
implies slightly reduced friction either with respect to the rough surface
or between individual grains. The reduction could have been caused by  the
elimination of non-spherical dust particles owing to washing the sample.
The four samples of copper are nice examples of the effect of particle shape.
A systematic increase of $\theta_1$ and $\theta_r$ detected by changing
shape anisotropy in the order of spherical beads ($\eta=0.63$), particles
with irregular but rounded shapes ($\eta=0.5$) and the two sets of particles
with very anisotropic dendritic shapes ($\eta=0.25$ and $\eta=0.33$).

The influence of the boundary conditions can have a profound effect on the
conditions of the granular flow. The usual no-slip boundary condition 
appropriate for a fluid is probably never completely satisfied for a granular 
flow and certainly depends on surface roughness.  Further, the role of the 
surface in damping energy is only recently beginning to attract attention 
\cite{he2007pc} and has not been considered in the context of granular flows 
on an incline.  Thus, it is important to evaluate the dependence of our
results on surface roughness and, in principle, on surface restitution 
coefficient. Although we do not consider the latter here, the systematic 
of a flow rule comparison may depend on the damping properties of the 
surface which may help explain differences between flow on a soft felt 
surface, on a glass plate with glued on hard particles, or a hard 
surface covered with sandpaper.

\begin{figure}[h]
\resizebox{80mm}{!}{
\includegraphics{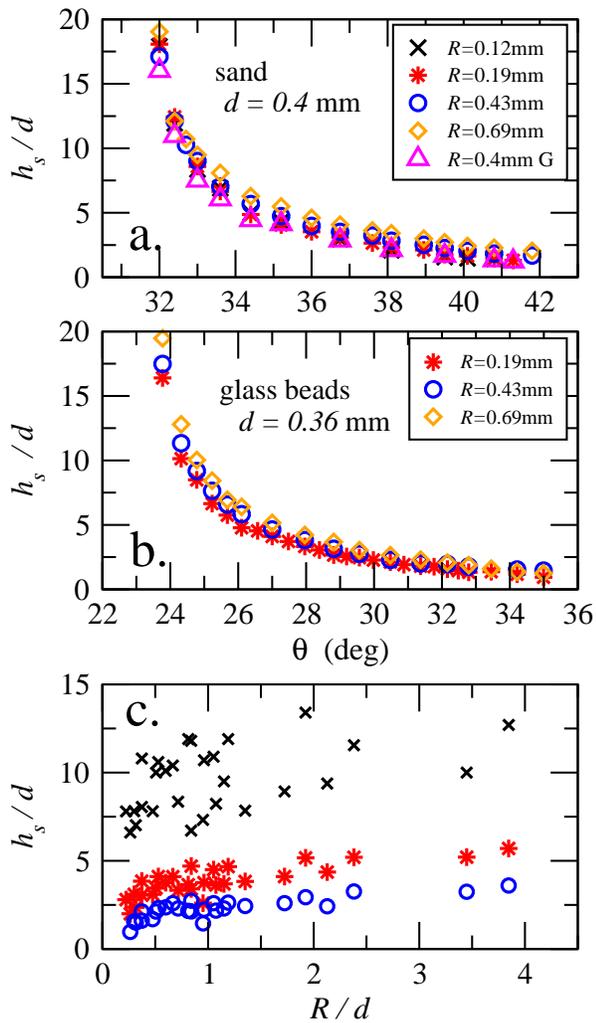}
}
\caption{
(Color online) $h_s$ vs. $\theta$ for a) sand with $d=0.4$ mm and b) glass 
beads with $d=0.36$ mm for various values of surface roughness: surface covered
by sandpaper with $R=0.12$ mm (x), $R=0.19$ mm ($\star$), $R=0.43$ mm ($\circ$)
and $R=0.69$ ($\diamond$), surface covered by one layer of $d=0.4$ mm sand 
particles ($\triangle$).  c) $h_s/d$ as a function of $R/d$ for 
$\tan \theta / \tan\theta_r = 1.1$ (x); $1.25$ ($\star$); and $1.4$ ($\circ$)
for sand and glass beads.
}
\label{surface}
\end{figure}

To study the dependence of our results on surface roughness, we measured
the dependence of $h_s/d$ on plane roughness $R$, shown in Figs.\ 
\ref{surface}(a) and \ref{surface}(b) for sand with $d=0.4$ mm  and glass 
beads with $d=0.36$ mm on four different sandpapers with nominal roughness
of $R=0.12$ mm, $R=0.19$ mm, $R=0.43$ mm and $R=0.69$ mm (grits 120, 80, 40 
and 24 respectively). For sand $h_s/d$ was also determined on a surface 
prepared by gluing one layer of the same grains onto the plate. For both sand
and glass beads, Figs.\ \ref{surface}(a) and \ref{surface}(b), a slight
increase of $h_s/d$ is observed with increasing plane roughness. For the 
case of sand with $d=0.4$ mm the curve measured on the surface prepared by
gluing the same grains was the most similar to the curves taken on sandpaper
with $R=0.12$ mm or $R=0.19$ mm, {\it i.e.}, the surface friction for 
sandpaper is somewhat larger than when the surface is covered with sand 
glued to the surface.

We determine the relative effect of surface roughness on the determination 
of $h_s$ by comparing data for sand and glass beads   for different values of
$d$ and $R$.  The value of $h_s/d$ increases as a function of $R/d$ as shown
for three values of $\tan \theta / \tan \theta_r$ in Fig.\ \ref{surface}(c).
At plane inclinations close to the bulk angle of repose $\theta_r$, the curve
seems to saturate [see the curve taken at $\tan \theta / \tan \theta_r = 1.1$
in Fig.\ \ref{surface}(c)] but for larger plane inclinations, {\it i.e.,} 
for thinner layers, a slight increase of $h_s/d$ is observed over the measured 
range of $R/d$. The increasing tendency of $h_s/d$ indicates that the effective
friction near a rough surface increases slightly with increasing plane 
roughness. Near the rigid surface the particles have less freedom to rearrange
so that in order to shear the medium has to dilate more \cite{pore1996} 
yielding a larger effective friction, compared to the case
of the bulk material. The growing value of $h_s/d$ matches the overall tendency
of the data reported in \cite{gdrmidi2004} using monodisperse glass beads on
surfaces prepared by gluing one layer of glass beads on a plate. We did not,
however, find any significant height maximum corresponding to a particular
plane roughness reported in \cite{gdrmidi2004,goth2003}.
Note that a stronger difference in $h_s/d$ was detected when the values
measured on a solid rough surface (similar to our case) and on velvet cloth
were compared \cite{gdrmidi2004}. As discussed above, this may be more a result
of surface damping than surface roughness.

The homogeneous dense-flow regime existed for moderate plane inclinations
where $\tan \theta /\tan \theta_1$ was in the range $1.1-1.45$.
According to our measurements \cite{boec2006vacuum}, the density of the flow
in this regime was decreased slightly with increasing $\theta$ but was always
larger than $0.8\rho_s$ where $\rho_s$  is the close packed static density of
the material, in accordance with other experimental data \cite{po1999,fopo2003}
and with numerical simulations \cite{sier2001,sila2003,mina2005}.

\subsection{Flow rule for glass beads and sand}
\begin{figure}[h]
\resizebox{74mm}{!}{
\includegraphics{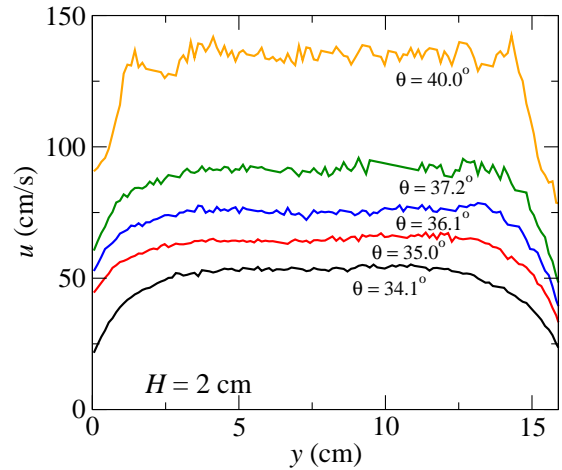}
}
\caption{
(Color online)
Transverse velocity profiles of the flow for sand with $d=0.4$ mm
at a hopper opening of $H=2$ cm.
}
\label{velo-profilok}
\end{figure}

We next present measurements of flow velocity obtained using the space-time
technique described above. 
To demonstrate that sidewall boundaries do not affect the velocity near the
channel center, we consider the transverse velocity profiles shown in
Fig.\ \ref{velo-profilok} for a hopper opening of $H=2$ cm for several values of
$\theta$. The data show that friction with the smooth sidewalls is much
less important than friction with
the rough bottom plate so that the sidewalls only produce a lateral boundary
layer at the edge of the channel with a characteristic thickness of 2-3 cm.
Over the remaining 80\% of the channel width, $u$ is very constant. For
determination of the flow rule, the velocity $u$ and thickness $h$
were measured at the channel center.

\begin{figure}[h]
\vspace*{0.1cm}
\resizebox{85mm}{!}{
\includegraphics{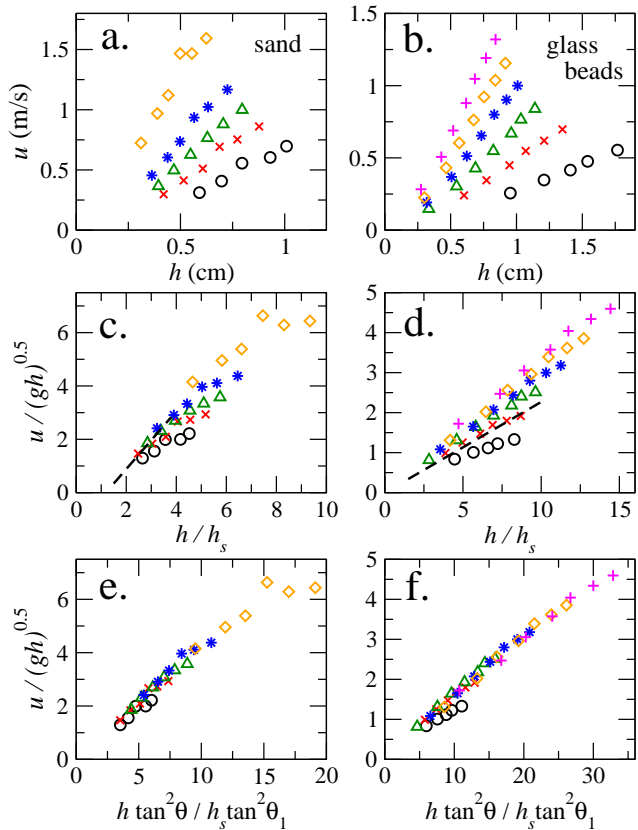}
}
\caption{
(Color online)
The flow velocity $u$ as a function of $h$ for a) sand
with $d=0.4$ mm and b) glass beads of $d=0.36$ mm.
Corresponding $\theta$ values of $\theta$  for a) sand
$34.1^\circ$ ($\circ$),
$35.0^\circ$ ($x$),
$36.1^\circ$ ($\bigtriangleup$),
$37.2^\circ$ ($\star$),
$40.0^\circ$ ($\diamond$);
b) for glass beads
$25.6^\circ$ ($\circ$),
$26.8^\circ$ ($x$),
$28.0^\circ$ ($\bigtriangleup$),
$29.4^\circ$ ($\star$),
$30.7^\circ$ ($\diamond$),
$32.0^\circ$ ($+$).
$u/\sqrt{gh}$ vs
$h/h_s$ for c) sand and d) glass beads. $u/\sqrt{gh}$  for e) sand and f) glass beads  vs. $h\tan^2 \theta/h_s \tan^2 \theta_1$.
}
\label{sand-glass-example}
\end{figure}

We now consider $u$ as a function of $h$ for sand and glass beads, presented
in a variety of forms to test both PFR
\cite{po1999} and PJFR \cite{je2006}.
In Figs.\ \ref{sand-glass-example}(a) and \ref{sand-glass-example}(b), we show $u$ as a
function of the flow thickness $h$ for sand with $d=0.4$ mm and glass beads
with $d=0.36$ mm.
In Figs.\ \ref{sand-glass-example}(c) and \ref{sand-glass-example}(d),
the same data are presented in dimensionless form according to the flow rule
$u/\sqrt{gh} \sim h/h_s$ suggested by Pouliquen \cite{po1999}.
For comparison, we include the curves measured by Pouliquen for glass
beads with $d=0.5$ mm and sand with $d=0.8$ mm \cite{fopo2003}, correcting
for the difference between depth averaged velocity $U$ and surface velocity
$u$ that assumes a Bagnold velocity profile for which $u=1.67U$.

Our data cover a wider range of $u$ and $h$ than previously measured,  partly
because of the smaller grain size, but also owing to the measurement technique.
Namely, measuring the surface velocity in the stationary regime was much more
straightforward for us than detecting the velocity of the front and thereby
determining $U$. For the detection of the front velocity the difficulty was 
that in contrast to the simple monotonic increase of the height at the flow 
front (reported in \cite{po1999}), in some cases and particularly for 
anisotropic grains we observed a larger height in the vicinity of the front.
In other cases, typically for larger (spherical) grains, the front was less 
defined with some grains rolling ahead of the front, {\it i.e.}, saltating.

The collapse of the data curves for sand and glass beads in
Figs.\ \ref{sand-glass-example}(c) and \ref{sand-glass-example}(d) is not perfect.
In these dimensionless units higher plane inclinations still result in somewhat
faster flow. We therefore consider the modified PJFR scaling
\cite{je2006} that includes a $\tan^2\theta$ correction to the
$h/h_s$ term. In Figs.\ \ref{sand-glass-example}(e) and \ref{sand-glass-example}(f),
we plot our data in terms of this modified scaling form, namely, $u/\sqrt{gh}$
versus $h\tan^2\theta /h_s\tan^2\theta_1$.
\begin{figure}[ht]
\vspace*{0.1cm}
\resizebox{80mm}{!}{
\includegraphics{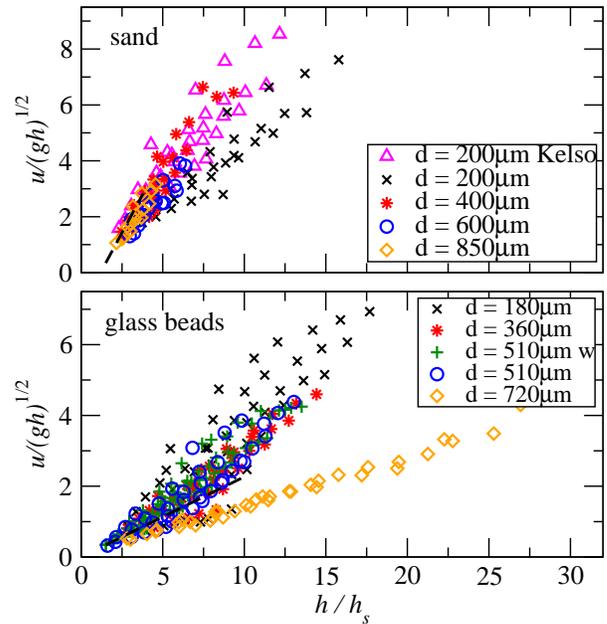}
}
\caption{
(Color online)
Dimensionless flow velocity $u/\sqrt{gh}$ vs $h/h_s$ for sand and glass beads
as a test of PFR. The grain
diameter $d$ is indicated and w designates the case of washed glass beads.
The dashed lines correspond to the velocity data taken for glass beads with
$d=0.5$ mm and sand with $d=0.8$ mm from \cite{fopo2003}.
}
\label{height-velo}
\end{figure}
The PJFR produces improved scaling relative to PFR as demonstrated in
Figs.\ \ref{sand-glass-example}(c) and \ref{sand-glass-example}(d).
Another consequence of the theory is a prediction for the density decrease
with increasing $\theta$ as $\rho/\rho_s=1-B\cdot$tan$^6\theta$ (where 
$\rho_s$ stands for the close packed static density).  Our data for the 
mean density, reported elsewhere \cite{boec2006vacuum}, are well fit by 
the theoretical form with a value $B=0.52$.

We next extend our comparison of flow rules to the whole set of sand and glass
beads used in this study. The data taken for these materials were scaled in the
same manner as in the case of sand with $d=0.4$ mm and glass beads with 
$d=0.36$ mm as presented in Figs.\ \ref{sand-glass-example}(c) and 
\ref{sand-glass-example}(d), {\it i.e.}, using PFR. In Fig.\ \ref{height-velo},
we plot $u/\sqrt{gh}$ as a function of $h/h_s$ for sand and glass beads with a
variety of sizes.  The scatter of the data is a sign of an imperfect collapse 
for each material.
For comparison the flow rule measured by Pouliquen for sand and glass beads is
included (dashed lines) and agrees within the data scatter with our 
measurements.
\begin{figure}[ht]
\vspace*{0.1cm}
\resizebox{80mm}{!}{
\includegraphics{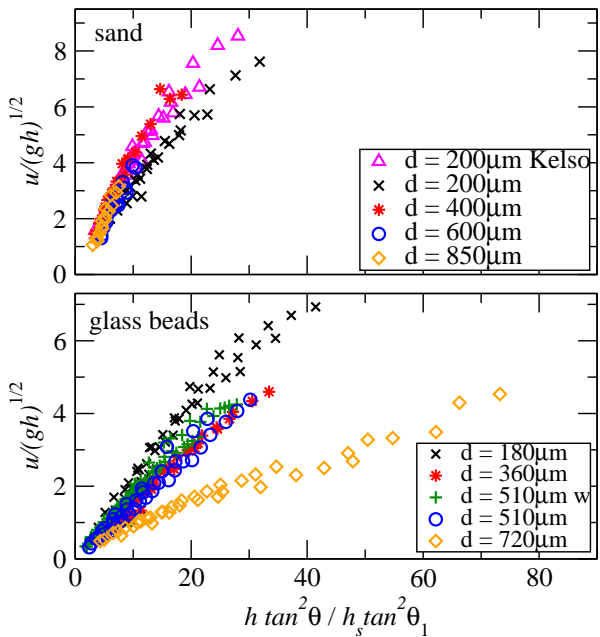}
}
\caption{
(Color online)
Dimensionless flow velocity $u/\sqrt{gh}$ vs $h \tan^2 \theta/ h_s\tan^2 \ \theta_1$
for sand and glass beads as a test of PJFR.
}
\label{height-velo-tan}
\end{figure}
Plotting $u/\sqrt{gh}$ as a function of $h\tan^2\theta /h_s\tan^2\theta_1$
for sand and glass beads yields an improved collapse, see
Figs.\ \ref{height-velo} and \ref{height-velo-tan}.

There are two things to notice about the curves in Figs.\ \ref{height-velo}
and \ref{height-velo-tan}, focusing more on the latter. First, there is the
linearity of the lines for different $d$. The glass bead data form quite nice
straight lines in support of a simple Bagnold rheology with $u \sim h^{3/2}$.
There is some remnant dependence on $d$ discussed below. The sand data are 
quite well collapsed and have a weaker variation on the grain size. The 
curves are, however, not straight lines but are slightly concave downward.
This deviation from linearity suggests a modification of the Bagnold rheology
is needed but the basic form captures the main details of the scaling. Second,
the nonzero offset $\gamma$ observed for sand for the case of the Pouliquen 
flow rule (Fig.  \ref{height-velo}) becomes approximately zero for the 
modified scaling relationship (see Fig. \ref{height-velo-tan}). This leads 
to a simpler quantitative comparison of these materials as the curves are
characterized by a single parameter $\beta$.

We now consider some of the details of the data from the perspective of the
particle size $d$.  Comparing the curves measured for various materials in
Fig.\ \ref{height-velo-tan}, we find curves with similar slopes for sand of
different sizes. Similarly, for glass beads the slopes $\beta$ of the curves
determined using PJFR and for samples of various $d$ do not differ much except
for the material with $d=0.72$ mm where $\beta$ is considerably smaller. This
difference cannot be quantitatively explained, but the case of the $d=0.72$ mm
glass beads could be special because the $R/d$ ratio is very small (0.26)
in this case.  For spherical beads there is a threshold value of $R/d$ below
which the beads simply roll down the plane. As we approach this threshold
by decreasing $R/d$ the value of $h_s/d$ drops rapidly. There is a stronger 
decrease of $h_s/d$, presumably resulting from the rolling effect of spherical
glass beads,  for spherical $d=0.72$ mm glass beads than for irregular 
$d=0.85$ mm sand particles as a function of decreasing $R/d$, obtained by 
varying the sandpaper roughness of $R=0.69$ mm, $0.43$ mm and $0.19$ mm.  
Using sandpaper with $R=0.12$ mm the $d=0.72$ mm beads already rolled down the
plane. The low value of $h_s/d$ could explain the low value of $\beta$ measured
for the glass beads of $d=0.72$ mm on sandpaper with $R=0.19$ mm. Generally,
the collapse suggests that the modified scaling theory describes the data quite
well provided the roughness ratio is larger than $R/d>0.3$ for glass beads and
$R/d>0.2$ for sand.

\begin{figure}[ht]
\vspace*{0.1cm}
\resizebox{75mm}{!}{
\includegraphics{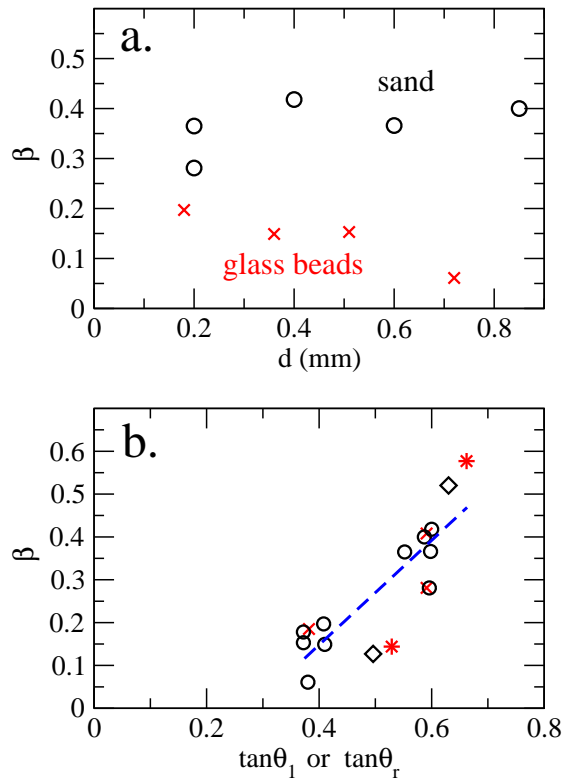}
}
\caption{
(Color online)
The PJFR slope $\beta$ vs a) grain size $d$ and b)
$\tan\theta_r$ (x) or $\tan\theta_1$ ($\circ$) for sand and glass bead samples.
The dashed line corresponds to a linear fit yielding
$\beta = 1.22\cdot \tan\theta_1 - 0.34$. Slopes for copper particles ($\star$
and $\diamond$) with $\eta = 0.33$ and $\eta = 0.50$ are included for 
comparison.
}
\label{flowruleslope}
\end{figure}

If the flow rule provided perfect collapse of the data, there would be no
residual dependence of the PJFR slope $\beta$ on $d$. This appears to be the case
for the sand flows where $\beta \approx 0.37$ independent of $d$ as illustrated
in Fig.\ \ref{flowruleslope}(a). On the other hand, the values of $\beta$ for
glass beads show a systematic decrease with increasing $d$. It is interesting
that sand with its somewhat anisotropic grains is less sensitive to size 
variation than the more idealized glass spheres. Again the rolling effect may 
play an important role here.

We now consider the behavior of the sand/glass-bead materials by plotting the
slope $\beta$ of the modified flow rule
$u/\sqrt{gh} = \beta \cdot h\tan^2\theta /h_s\tan^2\theta_1$
as a function of $\tan\theta_r$ or $\tan\theta_1$, see Fig.\ \ref{flowruleslope}(b).
A significant increase in $\beta$ is observed with increasing $\tan\theta_r$ or
$\tan\theta_1$. In a certain sense, these angles measure the degree of 
frictional interactions of the grains. This finding is in general agreement 
with earlier more limited data \cite{po1999,fopo2003} and gives a general 
characterization of the materials. Although we do not have enough data to 
unambiguously determine a functional dependence of $\beta$ on $\tan \theta_1$
(or $\tan \theta_r$), a linear fit to the data yields the relationship 
$\beta = 1.22\cdot \tan\theta_1 - 0.34$.

\subsection{Flow rule for copper particles}

The application of the flow rule scaling to the copper materials is 
an interesting
extension beyond those materials measured previously \cite{po1999,fopo2003}.
\begin{figure*}[ht]
\vspace*{0.1cm}
\resizebox{180mm}{!}{
\includegraphics{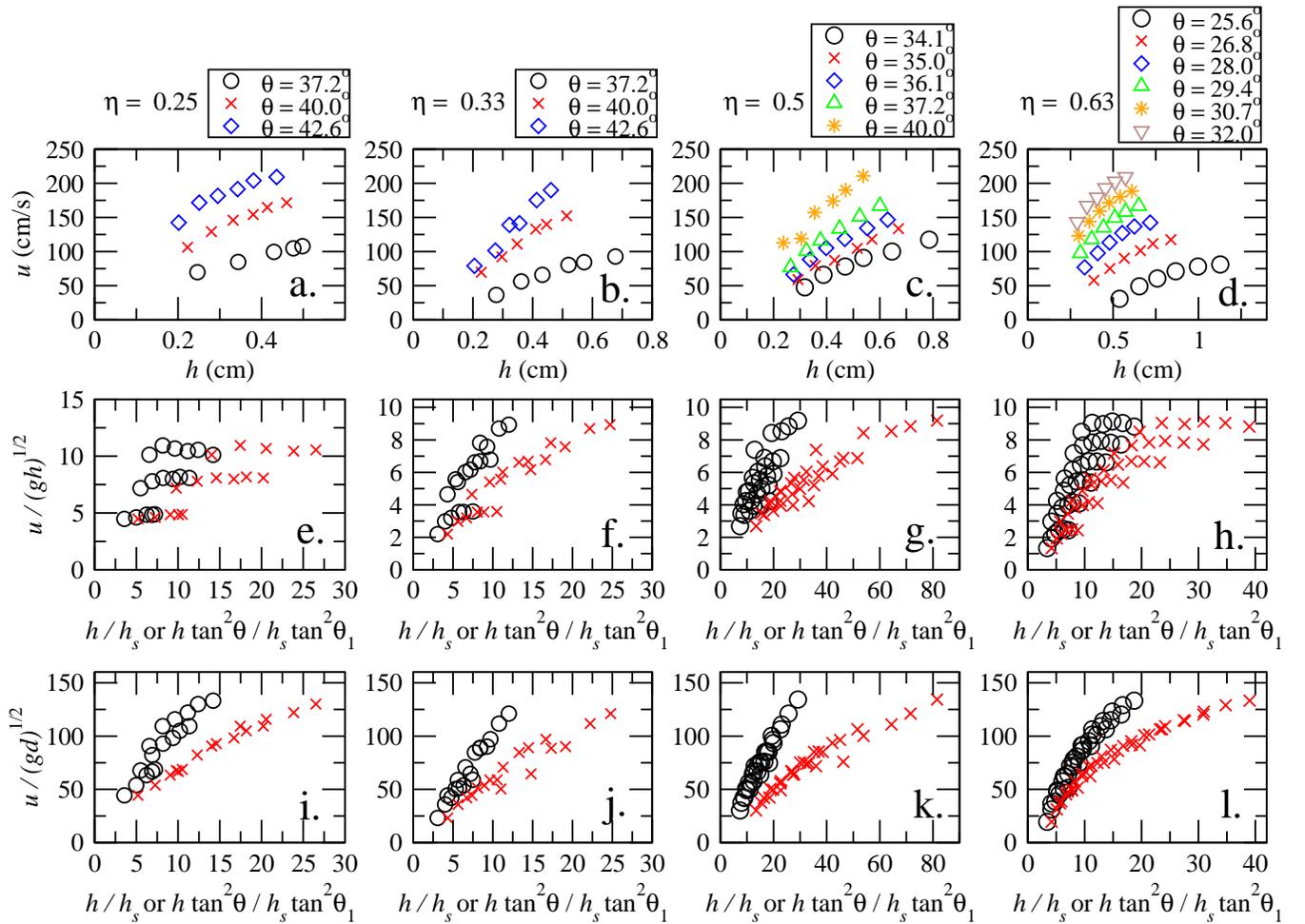}
}
\caption{
(Color online)
The flow velocity $u$ as a function of $h$ for the four set of copper
particles with  a) $\eta = 0.25$, b) $\eta = 0.33$, c) $\eta = 0.5$ and d) $\eta = 0.63$.
e)-h) $u/\sqrt{gh}$ vs  $h / h_s$ ($\circ$) and
$h \tan^2 \theta/ h_s\tan^2 \ \theta_1$ (x) for same $\eta$.
i)-l) $u/\sqrt{gd}$ vs  $h / h_s$ ($\circ$) and
$h \tan^2 \theta/ h_s\tan^2 \ \theta_1$ (x) for same $\eta$, i)-l).
}
\label{height-velo-copper}
\end{figure*}
In particular, the copper grains are metallic and thus not affected by static
charging. Further, the grains may oxidize producing different frictional
contacts than for the more inert sand and glass materials. Finally, the very
unusual shape anisotropy adds an additional level of complexity to the scaling
problem beyond the unknown differences in shape between the sand and glass 
beads. We proceed in the same way for the different copper grains as with 
the sand/glass-bead materials in that first we show the raw data for $u$ 
as a function of $h$ in Figs.\ \ref{height-velo-copper}(a) and 
\ref{height-velo-copper}(d). The data vary smoothly with $h$ for different 
values of $\theta$. Applying PFR or PJFR scaling as shown in 
Figs.\ \ref{height-velo-copper}(e)-\ref{height-velo-copper}(h) demonstrates
that either scaling does not work for the copper materials and is especially
poor for the spherical copper grains with $\eta = 0.63$. The apparent origin of
this poor collapse seems to be the assumed $h^{3/2}$ scaling implied 
by a Bagnold vertical velocity profile. If instead of dividing by $\sqrt{gh}$, 
one simply plots $u/\sqrt{gd}$ versus $h/h_s$ or the modified form 
$h \tan^2\theta /h_s\tan^2\theta_1$, the curves are now approximately 
collapsed, see Figs.\ \ref{height-velo-copper}(i)-\ref{height-velo-copper}(l).

In understanding this unexpected result, we first consider the spherical 
copper particles with $\eta = 0.63$ for which the comparison with the sand 
and glass beads might be thought to be most similar. The first thing to 
note is that there is a distinct concave downward curvature to the raw 
$u$ versus $h$ curves in Fig.\ \ref{height-velo-copper}(d) when compared 
to the case of sand or glass beads. Also, the character of the scaled 
curves is strongly non-linear for the case of copper with $\eta = 0.63$ 
and $\eta = 0.5$ (Figs. \ref{height-velo-copper}(g) and 
\ref{height-velo-copper}(h). Although we do not have a quantitative 
explanation for the behavior of the copper particle rheology, we note 
some ideas worth exploring. One issue of possible relevance is that the 
coefficient of restitution of soft metal particles, {\it i.e.}, brass,
copper, decreases with increasing velocity \cite{kuko1987,sthr2005} (the 
restitution coefficient for brass, which is harder than copper, decreases 
by about 8\% over the range of velocities in the present experiment - 
0-2.5 m/s) whereas the harder glass-bead/sand materials have a larger, 
velocity independent restitution coefficient. A velocity dependent 
(decreasing) restitution coefficient would lead to higher dissipation at 
larger velocities, but this effect has not been quantitatively studied. 
A recent study on soft particles with constant restitution coefficient 
\cite{sigr2007} suggests that the presence of long-lived contacts leads
to a modified rheology with a new term (similar to a Newtonian fluid), 
{\it i.e.}, $\sigma_{xz} =A \dot{\gamma}^2 +B\dot{\gamma}$. Such a 
relationship would lead to a faster growth of $u$ with increasing $h$ 
than $u \sim h^{3/2}$, a result that would lead to worse agreement for our 
copper data than did the Bagnold scaling.  Part of the issue here is the 
indirect measure of the bulk rheology provided by comparing the dependence 
of $u$ on $h$.

Another possible issue is the nature of the boundary condition for copper 
particles on the sandpaper surface. Unlike a fluid, a granular material can 
have a finite slip velocity at the surface. This finite velocity would 
complicate the scaling procedure and perhaps lead to spurious conclusions. 
Copper particles move somewhat faster for a given thickness $h$ owing to 
their smaller size and thus may develop a larger slip velocity. For example,
the copper particles have maximum velocities of order 2.2 m/s compared to
1.3-1.5 m/s for sand or glass beads over the same range of angle-corrected $h$.

\begin{figure}[ht]
\vspace*{0.1cm}
\resizebox{80mm}{!}{
\includegraphics{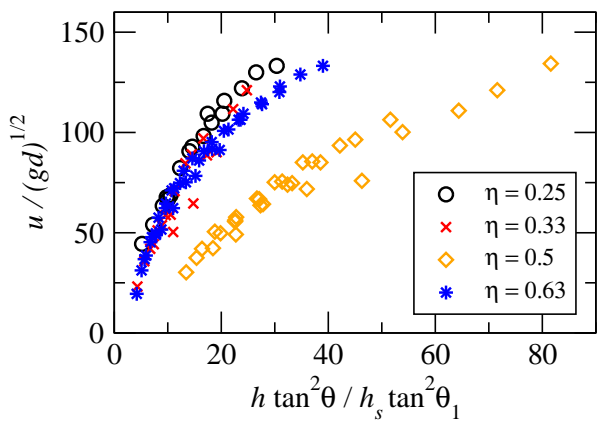}
}
\caption{
(Color online)
The dimensionless flow velocity $u/\sqrt{gd}$ as a function of the modified
dimensionless flow thickness $h \tan^2 \theta/ h_s\tan^2 \ \theta_1$ for the
four set of copper particles with  $\eta = 0.25$, $\eta = 0.33$, $\eta = 0.5$
and $\eta = 0.63$.
}
\label{height-velo-allcopper-unified}
\end{figure}

The other copper particles present a more complex situation.  First, we plot 
the $u/\sqrt{gd}$ versus the different angle-corrected scalings in 
Fig.\ \ref{height-velo-allcopper-unified}.  The data with $\eta = 0.25, 0.33$,
and $0.63$ collapse rather well but the curve with $\eta = 0.5$ has a quite 
different slope, about half of the other curves.  The difference in slope 
does not come from higher velocity but rather from larger $h$ relative to $h_s$.
In other words larger $h/h_s$ was needed for the realization of the stationary
flow regime, which results from a relatively larger dynamic friction 
coefficient, the source of which may be surface oxidation of the copper 
particles. This results in a lower value for $\beta$. This set of copper is 
also particular in that it is the only copper sample emitting strong sound 
during shearing, similar to but much stronger than the Kelso sand.

Although all of the copper particle data are collapsed better by not 
scaling $u$ by $\sqrt{gh}$, PJFR is not so bad for the $\eta = 0.33$
and $\eta = 0.50$ copper particles. Extracting a slope $\beta$ for those values
of $\eta$ yields curves that are consistent with the sand/glass-bead scaling
as a function of $\tan \theta_1$, see Fig.\ \ref{flowruleslope}.  Thus, even
though the copper particles are quite different, they still seem to show
the same qualitative dependence on $\tan \theta_1$ as the glass-bead/sand particles.

\section{Conclusions}
\label{conclude}
The most important findings of this work can be summarized as follows.
The surface velocity $u$ as a function of flow thickness $h$ of a granular
flow on a rough inclined plane was measured for fourteen different materials
in the dense, stationary flow regime.
All configurations were characterized by measuring the value of $h_s$
(the thickness of the layer remaining on the plane after the flow subsided)
as a function of the plane inclination $\theta$. The value of $h_s/d$ for sand
and glass beads increased slightly with increasing ratio of plane roughness
and grain diameter $R/d$ measured for four different values of $R$.
The $u(h)$ curves for sand and glass beads measured at various $\theta$ did not
perfectly collapse using the scaling law  $u/\sqrt{gh} \sim h/h_s$ proposed by
Pouliquen \cite{po1999}. An improved collapse was obtained using PJFR
 $u/\sqrt{gh} = \beta \cdot h\tan^2\theta /h_s\tan^2\theta_1$
where the factor $\tan^2\theta$ was suggested by a recent theory by Jenkins
\cite{je2006}.  For the sand/glass-bead materials, the PJFR slope $\beta$ 
 increases strongly with $\tan\theta_1$  yielding a
quantitative description of various materials, thereby extending our tools for
a better characterization and prediction of complex dynamical phenomena, such
as waves \cite{fopo2003} or avalanche propagation \cite{boha2005}.

Our results demonstrate that when the surface velocity is used to determine 
the flow rule, the PJFR scaling is superior to the earlier PFR approach.
For the original data set of Pouliquen \cite{po1999}, it is hard to determine 
which scaling form is better. Two possibilities are suggested. First, the 
uncertainty (mainly in $h_s$) in the original measurements does not allow a 
definitive comparison.  Second, measuring the depth averaged velocity at the
front is substantially different from measuring the surface velocity. If the
latter is correct then something unexpected is happening in the layer because
Bagnold scaling (or any monotonic vertical velocity profile starting from 
zero velocity, for that matter) implies that the ratio of surface velocity 
to depth averaged velocity is a constant, and a constant factor would not 
change the flow rule.  Although we cannot definitively rule out some strange
behavior, the superior fit of PJFR, the elimination of the need for an offset
$\gamma$, and the consistency of these results with Bagnold scaling suggests
that the apparent discrepancy between our results and former scaling analysis
\cite{po1999} results from larger uncertainty in the previous measurements.

For copper grains of different shapes, neither the Pouliquen form nor the 
Jenkins modified scaling works well in collapsing data taken for a variety 
of values of $\theta$.  Although the angle correction works decently, the 
normalization of $u$ by $\sqrt{gh}$ produces poorer scaling.  This suggests 
that for the copper particles, a Bagnold form for the vertical velocity 
profile does not hold.  An important future extension of this work would be 
to directly measure the velocity as a function of vertical position for the
different materials to determine the velocity profiles. Measurements of this
type are being planned to test the conjectures based on phenomenological flow
rule comparisons.

Finally, one must conclude that although the rheology for sand and glass beads
seems rather robust and well fit by the Pouliquen/Jenkins form, this is no
guarantee that more general materials satisfy this scaling relationship.
The copper measurements are puzzling because one might have expected the
nearly spherical copper beads to produce results similar to the spherical
glass beads.  That the Bagnold form does not seem to apply for copper
grains of different shapes and especially for the spherical ones is quite
surprising and unexpected.  Experiments on other metallic particles would be
very helpful in determining the origins of this effect. Finally, a more direct
probe of the interior dynamics of granular flows seems essential for 
determining the bulk flow rheology for general granular media.

This work was funded by the US Department of Energy under Contracts W-7405-ENG \&
DE-AC52-06NA25396. The authors benefited from discussions with J. Jenkins and I. Aranson.
T.B. acknowledges support by the Bolyai J\'anos research program,
and the Hungarian Scientific Research Fund (Contract No. OTKA-F-060157).

\end{document}